%% file: ms_alex_v57.tex
\defigdir{./}
\def\emulateapj{}
\ifdefined\emulateapj
  \documentclass[12pt,preprint]{emulateapj}
\else
  \documentclass[12pt,preprint]{aastex}  
\fi
\usepackage{aas_macros}
\usepackage{amsmath}
\usepackage{placeins}

\usepackage{xspace} 
\usepackage{listings}
\usepackage[dvipsnames,svgnames]{xcolor} 
\usepackage{soul} 
\usepackage{graphicx}
\usepackage{lineno}
\usepackage[T1]{fontenc}
\usepackage[utf8]{inputenc}
\usepackage{lmodern}

\usepackage{natbib}
\usepackage{hyperref}
\hypersetup{    
  colorlinks      = true,
  linkcolor       = {black},
  linkbordercolor = {white},
  citecolor       = {black},
  citebordercolor = {white},
  urlcolor        = {blue},
  urlbordercolor  = {white},
}

\ifdefined\emulateapj
  
\fi

\ifdefined\figdir
  \graphicspath{{\figdir}}
\else
  \graphicspath{{../figs/}}
\fi

\input{commands.tex}

\begin{document}

\title{Searching for Dark Matter Annihilation in the Smith High-Velocity Cloud}

\author{
Alex~Drlica-Wagner\altaffilmark{1}, 
Germ\'an A. G\'omez-Vargas\altaffilmark{2,3}, 
John W. Hewitt\altaffilmark{4,5}, 
Tim~Linden\altaffilmark{6}, and
Luigi~Tibaldo\altaffilmark{7}
}
\altaffiltext{1}{Center for Particle Astrophysics, Fermi National Accelerator Laboratory, Batavia, IL 60510, USA}
\altaffiltext{2}{Departamento de Fis\'ica, Pontificia Universidad Cat\'olica de Chile, Avenida Vicu\~na Mackenna 4860, Santiago, Chile}
\altaffiltext{3}{Instituto Nazionale di Fisica Nucleare, Sez. Roma Tor Vergata, Via della Ricerca Scientifica I-00133, Roma, Italy}
\altaffiltext{4}{CRESST, University of Maryland, Baltimore County, Baltimore, MD 21250, USA}
\altaffiltext{5}{NASA Goddard Space Flight Center, Greenbelt, MD 20771, USA}
\altaffiltext{6}{The Kavli Institute for Cosmological Physics, University of Chicago, Chicago, IL 60637, USA}
\altaffiltext{7}{W. W. Hansen Experimental Physics Laboratory, Kavli Institute for Particle Astrophysics and Cosmology, Department of Physics and SLAC National Accelerator Laboratory, Stanford University, Stanford, CA 94305, USA}



\begin{abstract}
Recent observations suggest that some high-velocity clouds may be confined by massive dark matter halos. 
In particular, the proximity and proposed dark matter content of the Smith Cloud make it a tempting target for the indirect detection of dark matter annihilation. 
We argue that the Smith Cloud may be a better target than some Milky Way dwarf spheroidal satellite galaxies and use $\gamma$-ray observations from the \Fermi Large Area Telescope to search for a dark matter annihilation signal. 
No significant $\gamma$-ray excess is found coincident with the Smith Cloud, and we set strong limits on the dark matter annihilation cross section assuming a spatially extended dark matter profile consistent with dynamical modeling of the Smith Cloud. 
Notably, these limits exclude the canonical thermal relic cross section ($\roughly \relic$) for dark matter masses $\lesssim 30\GeV$ annihilating via the \bbbar or \tautau channels for certain assumptions of the dark matter density profile; however, uncertainties in the dark matter content of the Smith Cloud may significantly weaken these constraints. 
\end{abstract}

\section{Introduction}
\label{sec:introduction}

Many theories predict that dark matter is composed of weakly interacting massive particles (WIMPs), which may annihilate into standard model particles in regions of high dark matter density~\citep[\eg,][]{2005PhR...405..279B,2008JCAP...07..013B}. Since the conventional WIMP mass scale resides in the \GeV to \TeV range, one of the principal products of dark matter annihilation would be a flux of $\gamma$ rays. For this reason, $\gamma$-ray searches with the Large Area Telescope on-board the {\it Fermi Gamma-ray Space Telescope} \citep[\Fermi-LAT;][]{2009ApJ...697.1071A} have provided some of the strongest limits on dark matter annihilation. 

Searches for dark matter annihilation traditionally target regions where an appreciable dark matter density is observationally confirmed and regions where a large dark matter density is strongly motivated by theory.
The former category includes the Milky Way dwarf spheroidal satellite galaxies, where stellar velocity measurements can often constrain the dark matter content to within a factor of two \citep[\eg,][]{2010ApJ...712..147A, 2011PhRvL.107x1303G, 2011PhRvL.107x1302A, Ackermann:2013yva}.
The latter category includes the Galactic Center \citep[\eg,][]{2011PhLB..697..412H, 2011PhRvD..84l3005H, 2012PhRvD..86h3511A, 2013PhRvD..88h3521G, Gomez-Vargas:2013bea} and clusters of galaxies \citep[\eg,][]{Ackermann:2010rg, SanchezConde:2011ap, Ando:2012vu, Han:2012uw}, where the predicted dark matter signal strength is uncertain by several orders of magnitude.
It is essential to examine multiple targets when searching for dark matter annihilation, since other astrophysical backgrounds may mimic a dark matter signal in a specific region. 
Additionally, the detection of dark matter in multiple Galactic environments would provide a valuable check on the predictions of cosmological simulations by mapping the dark matter distribution on Galactic and sub-Galactic scales. 

High-velocity clouds (HVCs) are a unique class of nearby Galactic substructures which may host significant dark matter content. HVCs are detected as cold clouds of neutral hydrogen (\Hi), and are characterized by their large peculiar velocities, which are often incompatible with Galactic rotation~\citep{1997ARA&A..35..217W}. Multiple HVCs exhibit cometary morphologies, indicative of stripping by an external medium~\citep{2000A&A...357..120B}, and also have peculiar velocities which imply previous encounters with the Milky Way disk~\citep{1990ApJ...356..130M}. The origin of HVCs is currently unknown; however, several popular theories have arisen. A subset of clouds appears to be correlated with the Magellanic Stream, suggesting an origin linked to interactions between the Milky Way and the Magellanic Clouds~\citep{1974ApJ...190..291M}. Other models suggest that HVCs may be associated with globular clusters, the polar ring, or may be jettisoned from the Galactic nucleus. 
Yet another possibility is that HVCs may be accreted from intergalactic space~\citep{1997ARA&A..35..217W}. 

The dark matter content of HVCs is controversial. In models where HVCs are correlated with the Magellanic Stream or are jettisoned from galactic environments, the dark matter content of HVCs is likely to be negligible. However, \citet{1999ApJ...514..818B} argue that the spatial distribution of a sub-population of HVCs is consistent with the expected distribution of dark matter subhalos. These HVCs are expected to contain a substantial dark matter mass, potentially orders of magnitude larger than their \Hi mass. \citet{2000A&A...354..853B} argue that the sizes of observed \Hi clouds are not consistent with a mass profile consisting solely of the observed gas, and that the outer regions of the clouds would be unbound without a significant dark matter component. However, \citet{2012A&A...547A..43P} analyze the disruption of HVCs by the Galactic disk and conclude that the majority of HVCs do not contain a significant dark matter component. 
An analysis of compact \Hi clouds by \citet{2012ApJ...758...44S} find that the majority of HVCs are likely associated with stellar outflows or the Magellanic Stream, and thus possess little or no dark matter. \citet{2008MNRAS.390.1691W} note that the difference between the observed distribution of HVCs and the expected distribution of cold dark-matter subhalos could be corrected if gas far from the Galaxy is highly ionized, preventing the detection of distant HVCs through 21-cm line observations. Furthermore, they note that only a small sub-population of HVCs must be associated with cold dark matter halos to match the expectations from simulations~\citep{2004ApJ...609..482K}.
Thus, the lack of a significant dark matter component in some HVCs does not preclude the existence of a sub-population of HVCs possessing large dark matter content.

The Smith Cloud is one of the better-studied HVCs, owing to its substantial mass, relative proximity, and location near the Galactic plane~\citep{1963BAN....17..203S}. As a low-metallicity HVC~\citep{2009ApJ...703.1832H} lacking any clear association with the Magellanic Stream, the Smith Cloud is a plausible candidate to host a significant dark matter halo---though it is worth noting that low-metallicity HVCs could also arise from the tidal disruption of metal-poor galaxies~\citep{2012A&A...547A..43P}. The Smith Cloud has a striking cometary structure, indicative of a previous interaction with the Galactic disk.
The Smith Cloud resides $2.9 \pm 0.3 \kpc$ below the Galactic plane at a Galactocentric distance of $7.6 \pm 0.9 \kpc$ and a heliocentric distance of $12.4 \pm 1.3\kpc$. 
The \Hi content of the Smith Cloud has a projected size of ${>}\,3 \times 1\kpc$ and a current mass of $\roughly 10^6\Msolar$, with the brightest \Hi emission located at $l,b = 38 \fdg 67,-13 \fdg 41$~\citep{2008ApJ...679L..21L}. 
The three-dimensional motion of the Smith Cloud can be derived from its distance, morphology, and the positional dependence of its systemic velocity relative to the local standard of rest. The derived orbital parameters suggest that the Smith Cloud passed through the Galactic plane $\roughly 70\Myr$ ago and is predicted to cross the plane again in $\roughly 27\Myr$~\citep{2008ApJ...679L..21L}.
This observation is puzzling, since the relatively low \Hi mass density of the Smith Cloud suggests that it should have been disrupted by such an encounter.  
Specifically, the gaseous component has a weak self gravity, significantly lower than the ram pressure force from an interaction with the Galactic disk, and complete dispersion of the Smith Cloud should occur on timescales of less than $1\Myr$. 

Recently, \citet[hereafter \NBH]{2009ApJ...707.1642N} modeled the interaction of the Smith Cloud with the Galactic disk, employing the observed dynamical properties of the Smith Cloud as determined by \citet{2008ApJ...679L..21L}. 
By modeling the tidal disruption and ram-pressure stripping of the Smith Cloud during its collision with the Galactic plane, \NBH determined that an additional massive component (\ie, dark matter) is required for the survival of the gas cloud. 
To address uncertainties in the dark matter density profile, \NBH simulated situations where the dark matter density follows a Navarro-Frenk-White~\citep[NFW;][]{1996ApJ...462..563N}, Burkert~\citep{1995ApJ...447L..25B}, or Einasto~\citep{2008MNRAS.391.1685S} profile. 
To constrain uncertainties resulting from prior encounters of the Smith Cloud with the Galactic plane, \NBH considered scenarios in which the Smith Cloud has transited the Galactic plane only once, as well as scenarios where the gas distribution of the Smith Cloud has reached a steady state due to repeated dynamical interactions with the plane. 
In all cases, \NBH found that a significant dark matter component is required to prevent the disruption of the Smith Cloud. 
To gravitationally bind the Smith Cloud, \NBH calculated that the dark matter mass within $1\kpc$ of the cloud center must exceed $2 \times 10^8\Msolar$ prior to the most recent stripping event. 
Moreover, tidal stripping during the most recent interaction with the Galactic plane is expected to have decreased the dark matter mass by approximately a factor of two, compared to a factor of five decrease in the gas content. 
This indicates that the Smith Cloud is currently composed of nearly 99\% dark matter, comparable to the dark matter fraction in dwarf spheroidal galaxies.

In this work, we use \Fermi-LAT data to search for excess $\gamma$-ray emission coincident with the Smith Cloud, accounting for the $\gamma$-ray emission from nearby point-like sources, Galactic foregrounds, and extragalactic backgrounds. Conventional models of the Galactic $\gamma$-ray foreground include a component representing the emission from Galactic cosmic rays interacting with interstellar gas in the Smith Cloud. 
It is essential to remove this component from the foreground model before searching for residual $\gamma$-ray emission associated with the Smith Cloud. 
No substantial residual flux is observed coincident with the Smith Cloud, and we place strong constraints on the dark matter annihilation cross section in scenarios where the Smith Cloud is dynamically bound by a dark matter halo with the characteristics derived by \NBH. 
Interestingly, we note that in many regimes these limits are stronger than those obtained from observations of dwarf spheroidal galaxies due to the proximity of the Smith Cloud. 
However, we stress that the dark matter content of the Smith Cloud is far more uncertain than that of dwarf spheroidal galaxies. 

The outline of this paper is as follows. 
In Section~\ref{sec:darkmatter} we model the $\gamma$-ray emission from dark matter annihilation in the Smith Cloud. 
In Section~\ref{sec:diffuse} we summarize the procedure for creating a Galactic foreground model that excludes the Smith Cloud.
In Section~\ref{sec:data} we describe our analysis of the \Fermi-LAT data in the region of the Smith Cloud. 
In Section~\ref{sec:results} we set limits on the $\gamma$-ray flux associated with the Smith Cloud and constrain the dark matter annihilation cross section. 
Finally, in Section~\ref{sec:conclusions} we discuss the dependence of these limits on the dark matter content of the Smith Cloud, compare these limits with those derived from observations of dwarf spheroidal galaxies and the Galactic Center, and discuss how this analysis could be extended to a larger population of HVCs. 

\section{Dark Matter Models of the Smith Cloud}
\label{sec:darkmatter}

The integrated $\gamma$-ray flux at Earth, $\phi_s$ ($\photon \cm^{-2} \second^{-1}$), expected from dark matter annihilation in a density distribution, $\rho(\vect{r})$, is given by
\begin{equation}
\begin{aligned}
   \phi_s(\Delta\Omega) = 
   & \underbrace{ \frac{1}{4\pi} \frac{\sigmav}{2m_{\DM}^{2}}\int^{E_{\max}}_{E_{\min}}\frac{\text{d}N_{\gamma}}{\text{d}E_{\gamma}}\text{d}E_{\gamma}}_{\PhiPP} \\
   & \times
   \underbrace{\vphantom{\int_{E_{\min}}} \int_{\Delta\Omega}\Big\{\int_{\rm l.o.s.}\rho^{2}(\vect{r})\text{d}l\Big\}\text{d}\Omega '}_{\Jfactor}\,.
\end{aligned}
\label{eqn:annihilation}
\end{equation}
Here, the \PhiPP term depends on the particle physics properties of dark matter---\ie, the thermally-averaged annihilation cross section, \sigmav, the particle mass, $m_\DM$, and the differential $\gamma$-ray yield per annihilation, $\text{d}N_\gamma/\text{d}E_\gamma$, integrated over the experimental energy range from $E_{\min}$ to $E_{\max}$.
The \Jfactor is the line-of-sight integral through the dark matter distribution integrated over a solid angle, $\Delta\Omega$. 
Qualitatively, the \Jfactor encapsulates the spatial distribution of the dark matter signal, while \PhiPP sets its spectral character. 

There is significant uncertainty in the dark matter density profile of the Smith Cloud, thus we calculate the \Jfactor for each of the three dark matter halo profiles fit by \NBH. 
The parameters of each profile were derived from the total dark matter tidal mass, $\Mtidal$, required to confine gas in the Smith Cloud during its most recent interaction with the Galactic disk. These tidal masses are calculated independently by \NBH for each dark matter profile. 
We describe each of the dark matter density profiles in terms of a scale radius, $r_s$, and a scale density, $\rho_0$, as listed in~\tabref{jfactor}. 
The Einasto profile depends on an additional parameter $\alpha$ which is set to a value of $0.17$.
\begin{align}
 \rho(r) & = \frac{\rho_0 r_s^3}{ r ( r_s + r)^2 } & &  \mathrm{NFW} \label{eq:nfw} \\
 \rho(r) & = \frac{\rho_0 r_s^3}{ (r_s + r) ( r_s^2 + r^2) } & & \mathrm{Burkert} \label{eq:burkert}\\
 \rho(r) & = \rho_0 \exp{ \left\{ -\frac{2}{\alpha} \left[ \left(\frac{r}{r_s}\right)^\alpha - 1 \right] \right \} } & &  \mathrm{Einasto} \label{eq:einasto}
\end{align}
To avoid peripheral regions where tidal stripping may alter the dark matter density, we truncate our model of the $\gamma$-ray intensity profile $1\degree$ from the center of the Smith Cloud. 
To simplify comparisons with other dark matter annihilation targets (\ie, dwarf spheroidal galaxies), we compute the integrated \Jfactor from the Smith Cloud within this $1\degree$ radius (\tabref{jfactor}). 
This radius contains $\roughly 60\%$ of the total predicted $\gamma$-ray flux when cuspy NFW or Einasto profiles are assumed and $\roughly 10\%$ of the total predicted flux from the cored Burkert model.
Thus, this choice of radius yields a conservative estimate for the total \Jfactor of the Smith Cloud since the dark matter distribution may extend to several degrees.

\begin{deluxetable}{ l c c c c c c c c}
\tablecaption{Summary of Smith Cloud dark matter halo parameters.}
\tablehead{
\colhead{Profile} & \colhead{$r_s$}    & \colhead{$\rho_0$} & \colhead{$\Mtidal$} & \colhead{\Jfactor} \\
        & \colhead{$(\kpc)$} & \colhead{$(\Msolar \kpc^{-3})$} & \colhead{$(\Msolar)$} & \colhead{$(\GeV^2 \cm^{-5} \sr)$} }
\startdata
NFW     & $1.04$ & $3.7\times 10^7$  & $1.1 \times 10^8$ & $9.6 \times 10^{19}$\\
Burkert & $1.04$ & $3.7\times 10^7$  & $1.3 \times 10^8$ & $4.2 \times 10^{18}$\\
Einasto & $1.04$ & $9.2\times 10^6$  & $2.0 \times 10^8$ & $1.8 \times 10^{20}$\\
\enddata
\tablecomments{Integrated \Jfactors are calculated over a solid-angle cone with radius $1\degree$ ($\Delta\Omega \sim 9.6 \times 10^{-4} \sr$).}
\label{tab:jfactor}
\end{deluxetable}


\section{Galactic Foreground Modeling}
\label{sec:diffuse}
The observed foreground $\gamma$-ray emission from the region surrounding the Smith Cloud is dominated by $\pi^0$-decay emission produced from cosmic rays interacting with the atomic and molecular hydrogen gas of the Milky Way.\footnote{The $\gamma$-ray emission from inelastic hadronic interactions is composed of many processes, the most important of which being the production of $\pi^0$, which decay primarily to $\gamma\gamma$.} 
The \GALPROP cosmic-ray propagation code can be used to model the diffuse Galactic $\gamma$-ray foreground from processes such as inelastic hadronic collisions, bremsstrahlung, and inverse-Compton scattering.\footnote{\url{http://galprop.stanford.edu}}
\GALPROP accounts for effects such as diffusion, reacceleration, and energy loss via mechanisms such as synchrotron radiation~\citep{1998ApJ...509..212S, 2009arXiv0907.0559S}. 
The distribution of target material is derived from surveys of the $2.6\mm$ CO and $21\cm$ \Hi lines, supplemented with interstellar reddening maps from infrared observations of interstellar dust.
Notably, the official \Fermi-LAT model of Galactic diffuse emission recommended for discrete source analysis includes a $\gamma$-ray emission component associated with the Smith Cloud.%
\footnote{\url{http://fermi.gsfc.nasa.gov/ssc/data/access/lat/BackgroundModels.html}} 
We remove gas correlated with the Smith Cloud from our analysis for two reasons. 
First, the intensity and spectrum of cosmic rays are poorly constrained at the distance of the Smith Cloud, which leads to considerable uncertainty in the predicted $\gamma$-ray flux. 
Second, removing gas from the Smith Cloud eliminates a potentially degenerate emission component which may result in artificially strong limits on the dark matter annihilation rate within the cloud.

\begin{figure}
  \plotone{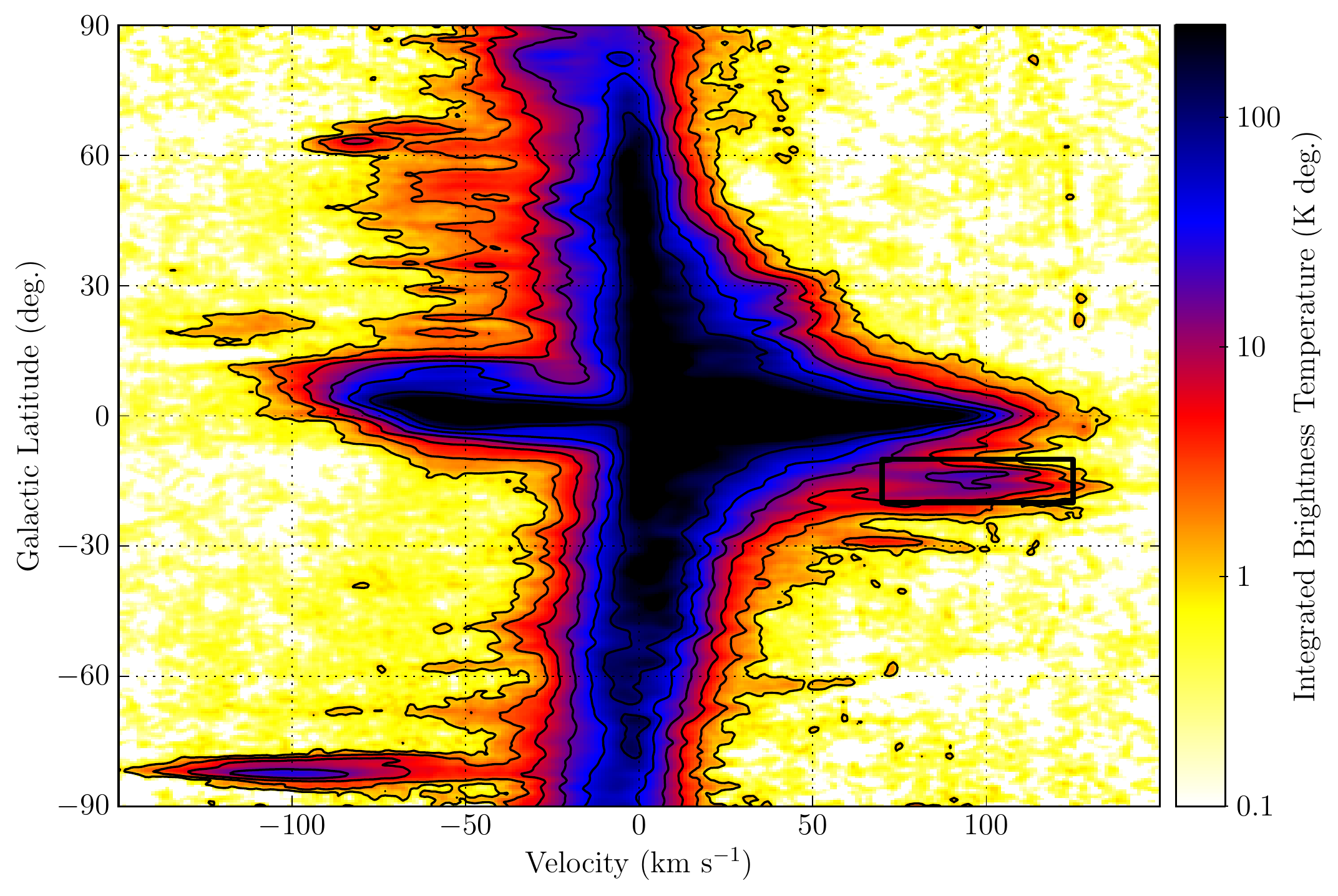}
  \caption{Latitude-velocity distribution of Galactic \Hi gas from the LAB survey \citep{Kalberla:2005ts} integrated over the longitude range of the Smith Cloud ($36\degree \le l \le 46\degree$). The color represents the integrated brightness temperature of the 21-cm \Hi line as a function of latitude and velocity with respect to the local standard of rest. Gas associated with the Smith Cloud is enclosed by the black box and is removed from our Galactic foreground model. \label{fig:LAB}}
\end{figure}

We create Galactocentric annuli for the \Hi gas distribution by transforming 21-cm brightness temperatures into column densities using the composite LAB survey \citep{Kalberla:2005ts} and the Galactic rotation curve given by \citet{clemens}. 
We follow the procedure employed by \citet{2012ApJ...750....3A} to excise the gas associated with the Smith Cloud from the Galactic gas distribution. 
Specifically, we remove gas in the region $36\degree \le l \le 46\degree$ and $-20\degree \le b \le -10\degree$, which has a velocity with respect to the local standard of rest in the range  70--125$\km \second^{-1}$ (Figure~\ref{fig:LAB}).%
\footnote{Gas with velocity ${>} 125\km\second^{-1}$ contributes less than $0.8\%$ of the total column density.} 
The primary uncertainty in the conversion from brightness temperature to column density comes from the assumed spin temperature ($T_S$) used to correct for the opacity of the 21-cm line. 
We find that the gas density in the region of the Smith Cloud changes by ${<}\,15\%$ when the assumed spin temperature is changed from $T_S=125 \Kelvin$ to $T_S=10^5 \Kelvin$ (\ie, the gas is optically thin).
When analyzing the $\gamma$-ray data we set $T_S=125 \Kelvin$ and find that this choice has little impact on our results. 
We follow the procedure of \citet{2012ApJ...750....3A} to trace the CO distribution from the 2-mm composite survey of \citet{Dame:2000sp}.
Due to the small CO content of HVCs~\citep{1999ApJ...523..163A}, we do not alter the CO map in the region of the Smith Cloud.
The Galactic foreground also contains a contribution from neutral gas that cannot be traced by the combination of \Hi and CO (so-called dark gas).
We follow the procedure of \citet{2012ApJ...750....3A} to trace the dark gas using the $E(B - V)$ reddening maps of \citet{Schlegel:1997yv}.
We incorporate a dark gas correction into the \Hi map after the Smith Cloud has been removed~\citep{2012ApJ...750....3A}.\footnote{The $E(B-V)$ correction excludes the Smith Cloud due to its low metallicity.}
We note that our procedure for removing the gas content of the Smith Cloud is very similar to the method used to remove gas associated with the Magellanic Clouds and M31~\citep[see Appendix B of][]{2012ApJ...750....3A}. 

These observations of the Milky Way gas profile supplemented by infrared observations of Galactic dust are input into the \GALPROP code to model the diffuse $\gamma$-ray emission corresponding to hadronic collisions, inverse-Compton scattering and bremsstrahlung radiation. 
To provide an accurate model for diffuse emission in the region of the Smith Cloud, we adopt the best-fit propagation parameters given by \citet{2011ApJ...729..106T}, specifically a convectionless diffusion constant of $8.32 \times 10^{28}\cm^2 \second^{-1}$ at a momentum of $4\GeV$, with a power-law momentum scaling $D(p) \propto p^{0.31}$, a height for the diffusion region of $5.4\kpc$, and an Alfv\'en velocity of $38.4\km\second^{-1}$. 
These parameters were inferred from a Bayesian analysis including the \Fermi-LAT data as an input, and the resulting model is well-fit to the Galactic diffuse $\gamma$-ray emission at intermediate latitudes corresponding to the Smith Cloud.
From this model we produce energy-dependent maps for the $\gamma$-ray emission from hadronic emission, bremsstrahlung, and inverse-Compton scattering. 
In principle, we would consider any alterations to the propagation parameters which are consistent with the local cosmic-ray primary-to-secondary ratios measured by satellite and balloon experiments. However, we find that this first attempt yields an accurate model of the observed diffuse $\gamma$-ray emission in the region of the Smith Cloud and no additional parameter-space scan is necessary. 

\ifdefined\emulateapj
  \vfill
  \eject
\fi

\section{Data Analysis}
\label{sec:data}
To search for excess $\gamma$-ray emission coincident with the Smith Cloud, we select a data sample corresponding to the first five years of \Fermi-LAT operation (2008 August 4 to 2013 August 4). 
We select events from the \code{P7REP} \code{CLEAN} class in the energy range from $500\MeV$ to $500\GeV$ and within a 15\degree radius of the Smith Cloud ($l,b = 38 \fdg 67,-13 \fdg 41$). 
Extending this analysis to lower energies would translate to a minor improvement in the sensitivity to low-mass dark matter models; however, below $500\MeV$ the rapidly changing effective area results in a stronger dependence on the spectral model assumed for the Smith Cloud. 
To reduce $\gamma$-ray contamination from the Earth's limb, we reject events with zenith angles larger than 100\degree and events collected during time periods when the magnitude of the rocking angle of the \Fermi-LAT was greater than 52\degree.

\begin{figure*}
  \plotone{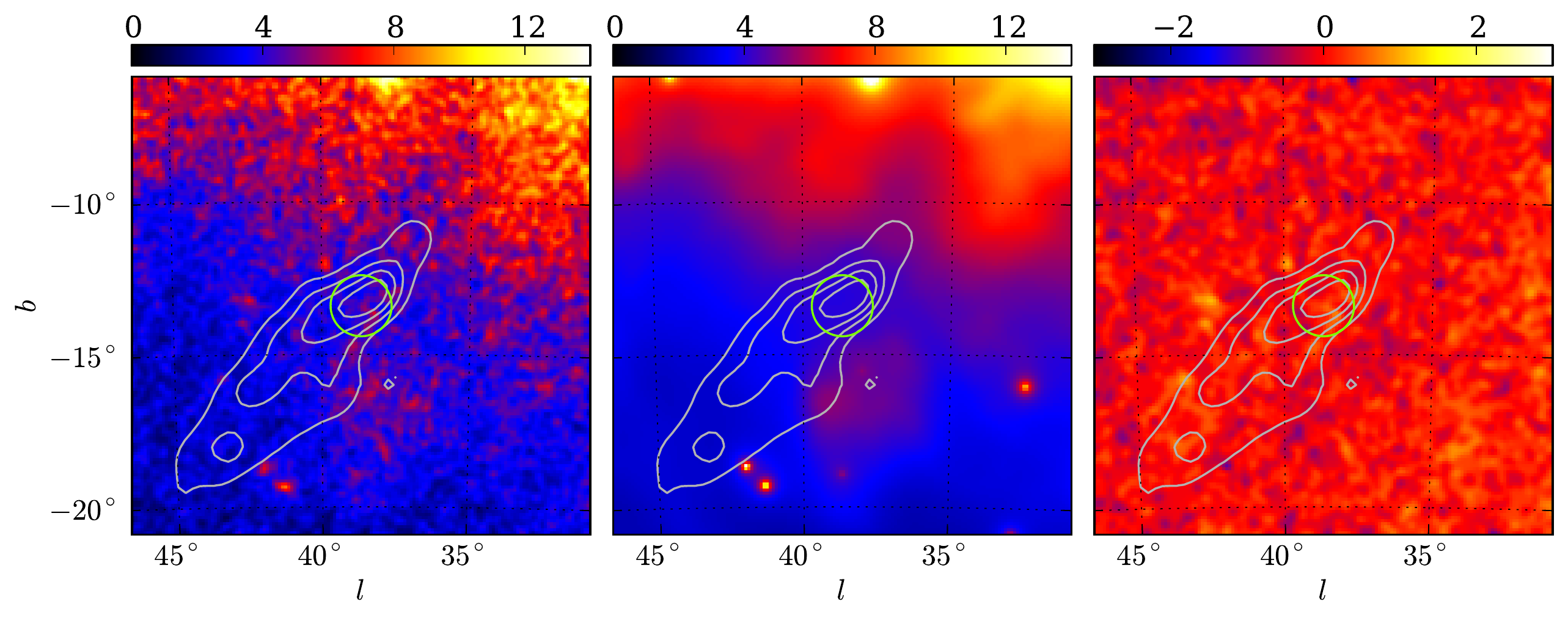}
  \caption{$15\degree \times 15\degree$ ROI surrounding the Smith Cloud in the energy range from $500\MeV$ to $500\GeV$. The gray contours represent the \Hi column density associated with the Smith Cloud ($1\times10^{20}\cm^{-2}  < \mathrm{N_{HI}} < 2.7 \times 10^{20}\cm^{-2}$), while the over-plotted circle shows the $1\degree$ truncation radius for the dark matter profile. Left: observed $\gamma$-ray counts map smoothed by a Gaussian kernel with standard deviation $0 \fdg 1$. Center: map of the background $\gamma$-ray emission model fit to the \Fermi-LAT data including diffuse and point-like backgrounds. Right: the Poisson probability of finding the observed number of counts in each pixel given the model prediction expressed as a Gaussian significance. \label{fig:spatial}}
\end{figure*}

We perform a binned maximum likelihood analysis of the $15\degree \times 15\degree$ region-of-interest (ROI) surrounding the Smith Cloud~(Figure~\ref{fig:spatial}).
We bin the \Fermi-LAT data in this ROI into 0 \fdg 1 pixels and 24 logarithmically-spaced bins of energy from $500\MeV$ to $500\GeV$.
We model the diffuse emission in this region using the templates for the hadronic, bremsstrahlung, and inverse-Compton emission derived in the previous section.
Because the hadronic and bremsstrahlung components are morphologically similar (both trace the interstellar gas in the Milky Way), we tie their relative normalizations in the $\gamma$-ray fit.
In addition to the diffuse Galactic foregrounds, the $\gamma$-ray data includes an isotropic contribution from extragalactic $\gamma$ rays and charged particle contamination. 
The spectrum of the isotropic $\gamma$-ray background is routinely derived from a high-latitude ($|b|\gtrsim10\degree$) fit to the \Fermi-LAT data, and is therefore dependent on the data selection and on the modeling of other $\gamma$-ray emission components (\ie, the Galactic diffuse emission).
It is difficult to derive the detailed spectrum of this component locally in the ROI of the Smith Cloud due to limited statistics at high energies and a morphological degeneracy with the inverse-Compton emission. Thus, we model the spectrum of the isotropic component with a broken power-law model which is simultaneously fit to the \Fermi-LAT data in the Smith Cloud ROI.
While a broken power-law model offers a reasonable fit to the Smith Cloud ROI, it does not capture the detailed energy dependence of the residual background.
To quantify the impact of this simplification we also perform the analysis with the standard isotropic background model, {\it iso\_clean\_v05.txt},%
\footnote{\url{http://fermi.gsfc.nasa.gov/ssc/data/access/lat/BackgroundModels.html}} and find that the results change by ${<}\,15\%$, which is much smaller than the fractional uncertainty in the dark matter distribution of the Smith Cloud.
In addition to the diffuse components, our model includes all sources from the second \Fermi-LAT source catalog within $20\degree$ of the Smith Cloud~\citep{2012ApJS..199...31N}. 
The flux normalizations of sources within $6\degree$ of the Smith Cloud are left free in the fit.

Since we are specifically interested in $\gamma$-ray emission associated with dark matter annihilation in the Smith Cloud, we model the Smith Cloud itself according to the spatially extended dark matter profiles described in \secref{darkmatter}.
We follow the prescription of~\citet{Ackermann:2013yva} to perform a bin-by-bin likelihood analysis of the $\gamma$-ray emission coincident with the Smith Cloud.
We first perform a single fit over the entire energy range to fix the normalizations of the diffuse and point-like background sources.%
\footnote{Fixing the normalizations of the background sources at their globally fit values avoids poor convergence in the fitting procedure resulting from the fine binning in energy and the degeneracy of the diffuse background components at high latitude.}
We then perform a likelihood scan over the normalization of the putative Smith Cloud dark matter source independently in each energy bin (this procedure is similar to that used to evaluate the spectral energy distribution of a source).
By analyzing each energy bin separately, we avoid selecting a single spectral shape to span the entire energy range at the expense of introducing additional parameters into the fit.
The common spectral model-dependent likelihood can be reconstructed by tying the signal normalization across the energy bins.
As a consequence, computing a single bin-by-bin likelihood function allows us to subsequently test many spectral models rapidly.
The bin-by-bin likelihood is additionally powerful because it presents the $\gamma$-ray data in a way that makes minimal assumptions about the $\gamma$-ray spectrum of the source of interest.
While the bin-by-bin likelihood function is essentially independent of spectral assumptions, it does depend on the spatial model of the Smith Cloud and must be derived for each profile in \tabref{jfactor}.

\section{Results}
\label{sec:results}

\begin{figure}
  \plotone{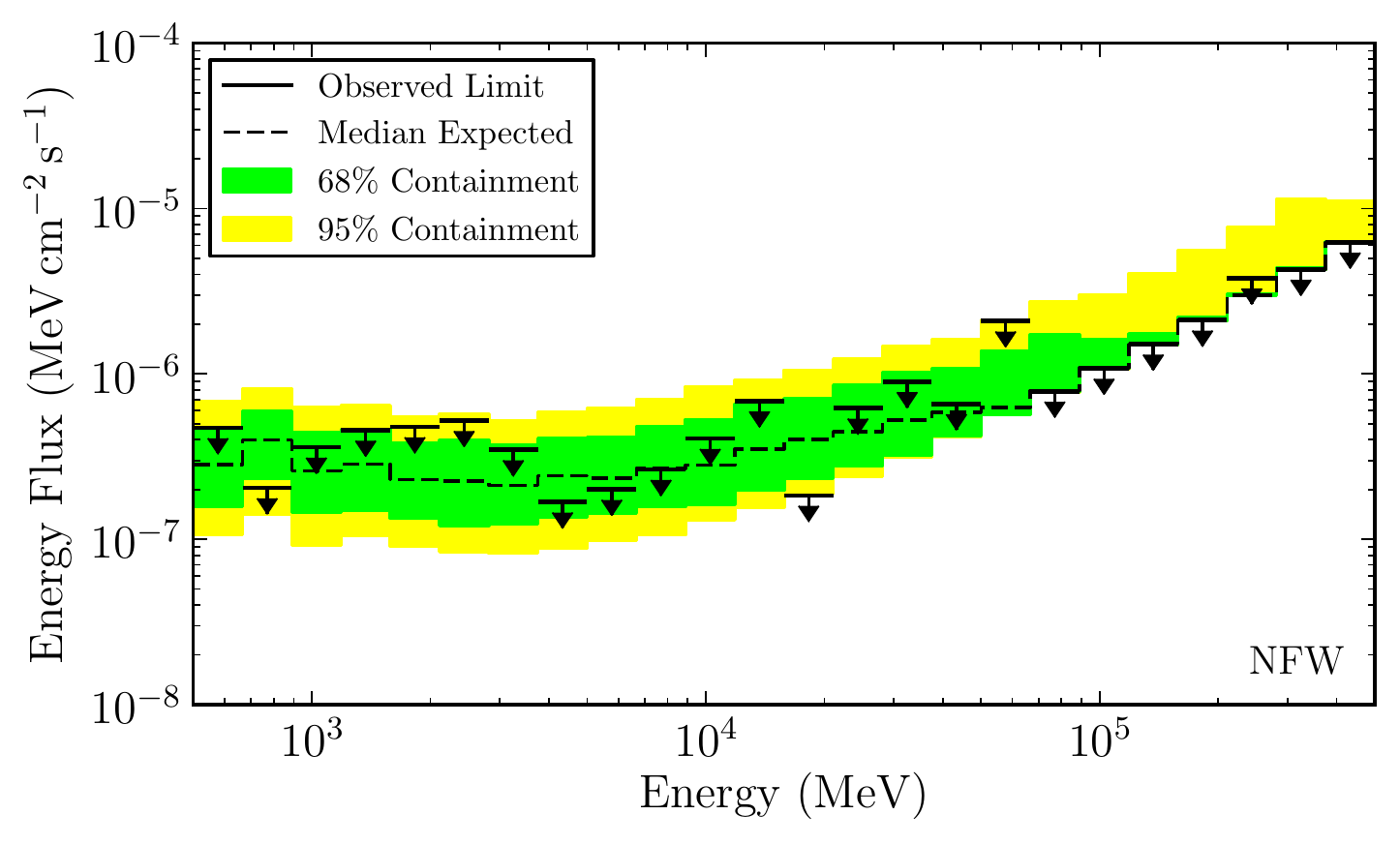}
  \caption{
Bin-by-bin energy-flux upper limits and expected sensitivities at 95\% CL for the Smith Cloud assuming an NFW dark matter profile. 
The 95\% CL upper limits derived from the data are shown by the black arrows.
The median sensitivity is shown by the dashed black line while the 68\% and 95\% containment regions are indicated by the shaded bands.\label{fig:sed_bands}
}
\end{figure}

\begin{figure}
  \plotone{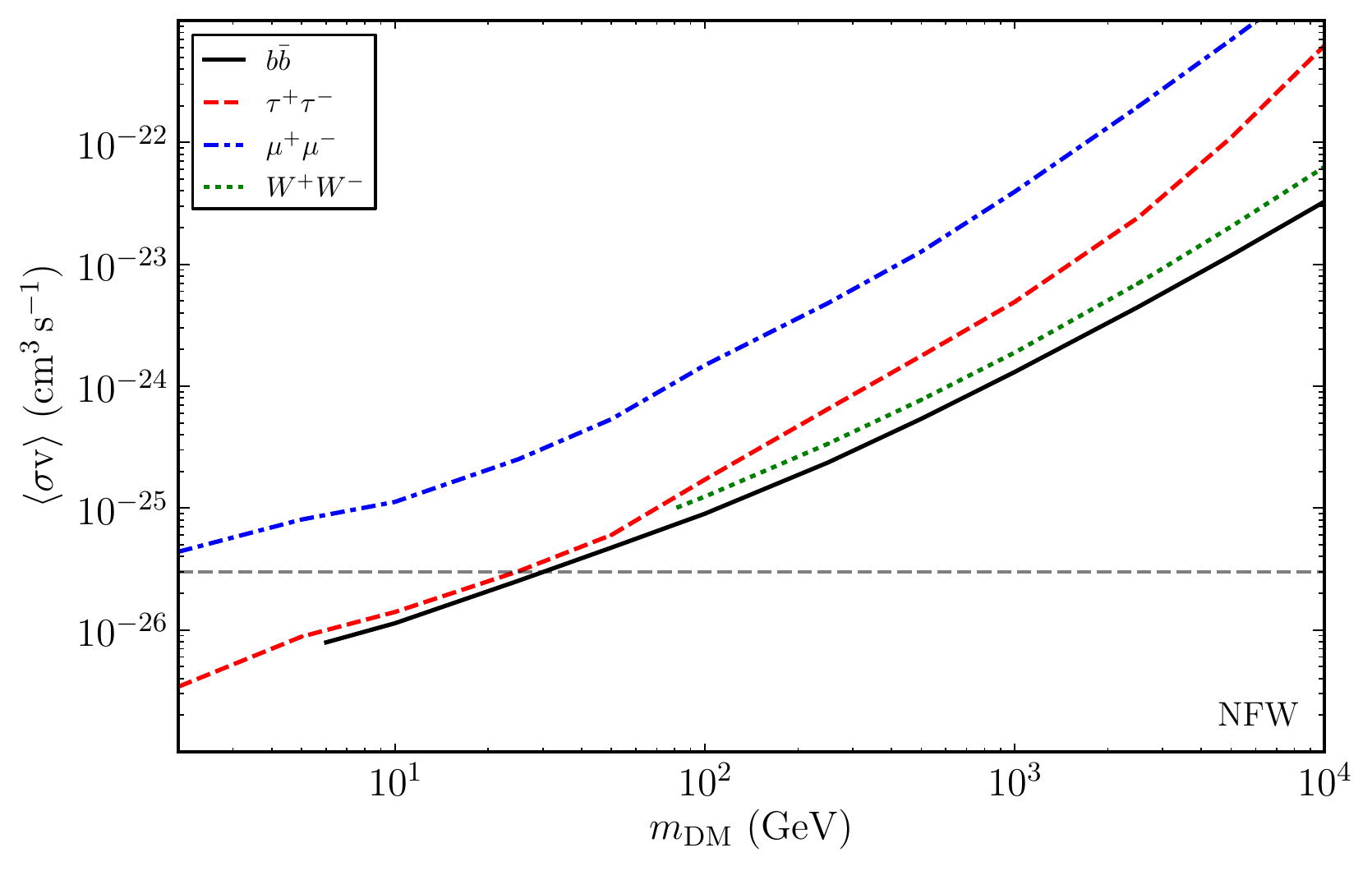}
  \caption{Upper limits at 95\% CL on the dark matter annihilation cross section as a function of the dark matter particle mass assuming the best-fit NFW profile from \tabref{jfactor} and annihilation through the \bbbar, \tautau, \mumu,  and \ww channels.  \label{fig:nfw_limits}}
\end{figure}

We find no statistically significant excess $\gamma$-ray emission coincident with the Smith Cloud in the energy range from 500\MeV to 500\GeV, and we set 95\% confidence level (CL) upper limits on the $\gamma$-ray flux within each energy bin (Figure~\ref{fig:sed_bands}).
These limits agree well with the expected sensitivity as calculated from 500 simulations of \Fermi-LAT instrument performance and the background $\gamma$-ray sources in the Smith Cloud ROI.
Under the assumption that the Smith Cloud is confined by a halo of dark matter, we utilize the absence of a $\gamma$-ray signal to set constraints on the dark matter annihilation cross section. Assuming an NFW profile with parameters listed in \tabref{jfactor}, we place constraints on the cross section for dark matter particles annihilating through the \bbbar, \tautau, \mumu, and \ww channels (Figure~\ref{fig:nfw_limits}).\footnote{Dark matter annihilation spectra were generated using DMFIT~\citep{Jeltema:2008hf,Ackermann:2013yva}.} Assuming an NFW profile, these constraints exclude the canonical thermal relic cross section ($\roughly \relic$) for dark matter masses $\lesssim 30\GeV$ annihilating via the \bbbar or \tautau channels. However, the \Jfactor calculated for the inner $1\degree$ of the Smith Cloud varies by more than an order of magnitude depending on the shape of the assumed dark matter profile. 
Current observations of the Smith Cloud offer no constraints on the shape of its dark matter profile and this uncertainty propagates directly into the constraints on the dark matter annihilation cross section (Figure~\ref{fig:bb_limits}). 
Uncertainty in the shape and content of the Smith Cloud dark matter halo dominates statistical uncertainties in the $\gamma$-ray data, systematic uncertainties in the modeling of the \Fermi-LAT instrument response, and systematic effects arising from the modeling of the diffuse $\gamma$-ray backgrounds.

The most significant deviation from the background-only hypothesis arises when fitting a dark matter particle with mass $5 \GeV$ annihilating to \tautau.  Incorporating this additional component increases the log-likelihood slightly, corresponding to a test statistic (\TS) of $\TS = 4.7$. This deviation is well below the conventional threshold ($\TS > 25$) for the detection of discrete $\gamma$-ray sources~\citep{2012ApJS..199...31N} and is spectrally consistent with hadronic emission produced from cosmic-ray interactions with the gas of the Smith Cloud. 
This hadronic component was explicitly removed from the astrophysical background model to provide a conservative constraint on the dark matter annihilation cross section. 
If the gas associated with the Smith Cloud is not removed from the astrophysical background model, the significance of the additional dark matter component is reduced to $\TS \approx 2$.  
The lack of significant cosmic-ray induced $\gamma$-ray emission from the Smith Cloud is not surprising due to the mass and distance of the cloud; however, a more detailed investigation is beyond the scope of this work.

\ifdefined\emulateapj
  \pagebreak
\fi
\section{Discussion and Conclusions}
\label{sec:conclusions}
Our analysis found no statistically significant $\gamma$-ray excess coincident with the Smith Cloud, allowing us to place limits on the dark matter annihilation cross section. 
Assuming the best-fit NFW profile from \NBH, these constraints are stronger than those derived from a four-year analysis of 15 stacked dwarf spheroidal galaxies for dark matter masses $\lesssim\,1\TeV$~\citep{Ackermann:2013yva}. The strength of these limits stems primarily from the fact that the Smith Cloud has a nominal \Jfactor approximately a factor of two higher than any of the dwarf spheroidal galaxies. Thus, we conclude that HVCs may be excellent indirect detection targets, motivating further investigation of the dynamics of these systems.  

\begin{figure}
  \plotone{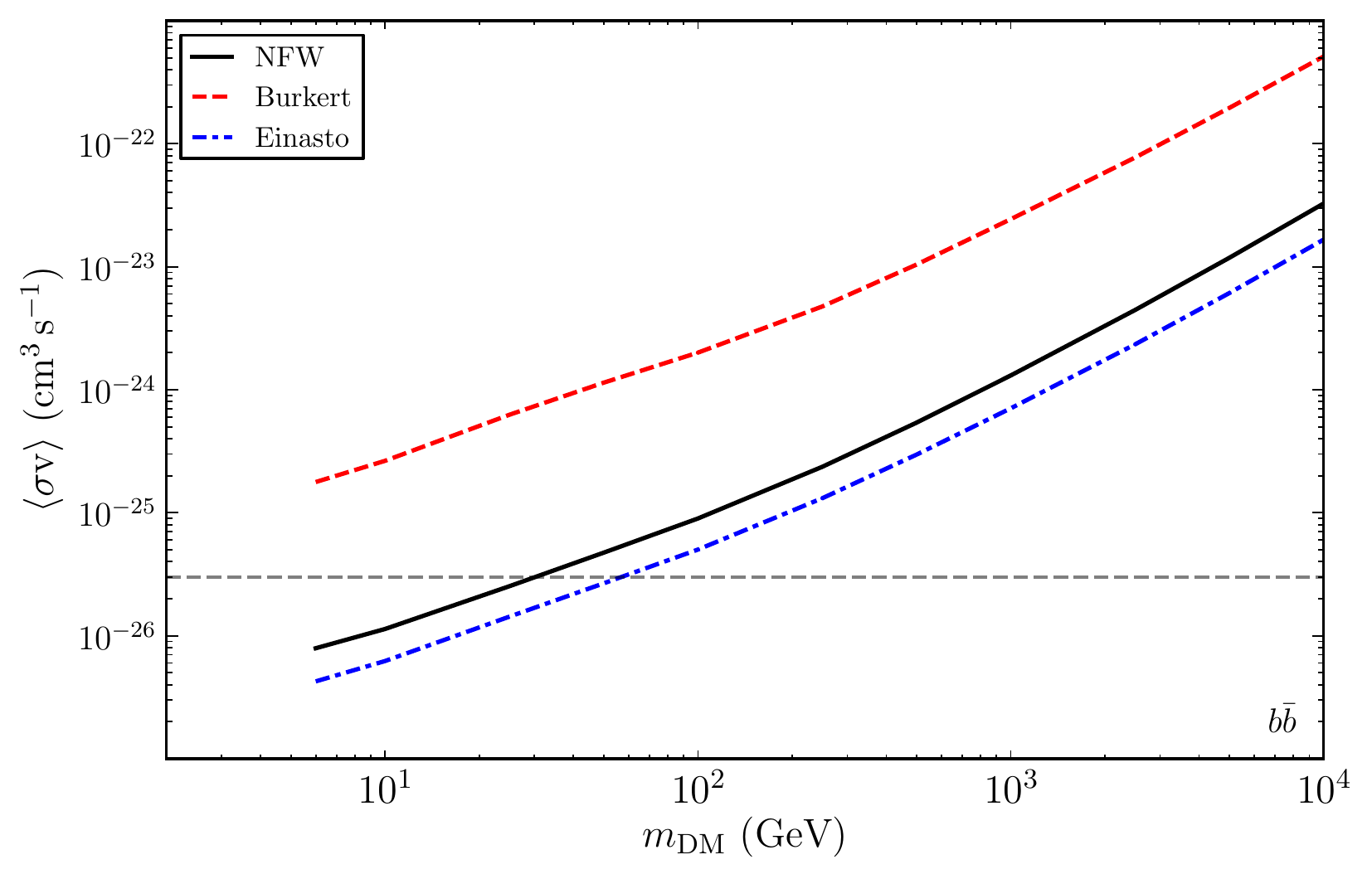}
  \caption{Upper limits at 95\% CL for the dark matter annihilation cross section to \bbbar as a function of the dark matter particle mass for different assumptions of the dark matter density profile. Each profile is normalized to the best-fit dark matter halo mass calculated by \NBH (\tabref{jfactor}). The annihilation signal is truncated at a radius of $1\degree$ from the assumed center of the Smith Cloud to mitigate possible impacts from tidal stripping. \label{fig:bb_limits}}
\end{figure}

It is important to note that while the constraints from dwarf spheroidal galaxies incorporate the uncertainty in the dark matter content of these objects, a similarly detailed understanding of the Smith Cloud is currently lacking. In Figure~\ref{fig:bb_limits}, we show that the limits on the dark matter annihilation cross section calculated under the assumption of a Burkert density profile are weaker than those calculated for the NFW and Einasto profiles by a factor of $\roughly 40$. 
It is likely that none of these analytic profiles realistically describes the dark matter profile over the full extent of the Smith Cloud, and thus an accurate description of the Smith Cloud density profile (including its overall normalization) currently remains the largest uncertainty in setting constraints on the dark matter annihilation cross section. 

We emphasize that identifying additional targets for indirect detection is particularly important in light of recent reports of excess $\gamma$-ray emission compared to current $\gamma$-ray diffuse emission models from the direction of the Galactic Center, which are possibly consistent with dark matter annihilation~\citep[\eg,][]{2013PhRvD..88h3521G,2014arXiv1402.4090A, 2014arXiv1402.6703D}. 
The cross section fit to the proposed Galactic Center excess barely evades the constraints derived from the Smith Cloud assuming an Einasto dark matter profile but lies a factor of $\roughly40$ below the limits derived assuming a Burkert profile. 
We find no statistically significant excess of $\gamma$-rays in the 1--3$\GeV$ region where the $\gamma$-ray excess from the Galactic Center is most pronounced. 
Even if a statistically significant $\gamma$-ray excess were to be found coincident with the Smith Cloud, a more thorough analyses of diffuse astrophysical $\gamma$-ray emission would be necessary in order to determine the origin of the emission. We emphasize that new indirect detection targets offer a compelling and complementary method to improve our sensitivity to dark matter annihilation using currently available data. 

The techniques developed in this paper can be readily extended to other Milky Way HVCs. 
More than 560 HVCs have been detected surrounding the Milky Way~\citep{1991A&A...250..509W}. 
While the distances to many HVCs are uncertain, these systems are relatively close to the solar position compared to the population of dwarf spheroidal galaxies. 
Future efforts to understand the dark matter content of HVCs and subsequently to perform a joint likelihood analysis of multiple HVCs may provide a sensitive test of dark matter annihilation.

We note that during the final preparation of this paper, \citet{2014arXiv1404.3209N} reported upper limits on the dark matter annihilation cross section from \Fermi-LAT observations of the Smith Cloud. 
While our results are qualitatively similar, we note three key differences between the \Fermi-LAT analyses in these studies. 
First, we remove gas correlated with the Smith Cloud before producing constraints on the Smith Cloud $\gamma$-ray emission, thus setting conservative constraints on the dark matter annihilation signal. 
Second, we utilize detailed spectral information when calculating upper limits on the dark matter annihilation cross section, rather than an integrated flux upper limit derived assuming a fixed power-law model for the $\gamma$-ray emission from the Smith Cloud. 
This allows us to more sensitively test specific dark matter masses and annihilation channels. 
Third, we model the dark matter content of the Smith Cloud as a spatially extended source with a distribution that extends out to $1\degree$, consistent with the profiles reported in \NBH. 
This last distinction significantly enhances the predicted dark matter annihilation signal from the Smith Cloud, and improves the constraints on the dark matter annihilation cross section by more than an order of magnitude (for an NFW profile) compared to the analysis of~\citet{2014arXiv1404.3209N}.

\section{Acknowledgments}
We would like to thank Luca Baldini, Seth Digel, Guðlaugur J\'ohannesson, and Miguel S\'anchez-Conde for helpful discussions. 
This project is partially supported by the NASA \Fermi Guest Investigator Program Cycle 6 No.~61330. TL is supported by NASA through Einstein Postdoctoral Award No.~PF3-140110. The work of GAGV was supported by Conicyt Anillo grant ACT1102 and the Spanish MICINN's Consolider-Ingenio 2010 Programme under grant MultiDark CSD2009-00064 and grant FPA2012-34694.
ADW received partial support from the Department of Energy Office of Science Graduate Fellowship Program (DOE SCGF) administered by ORISE-ORAU under Contract No. DE-AC05-06OR23100.

The \Fermi-LAT Collaboration acknowledges generous ongoing support from a number
of agencies and institutes that have supported both the development and the operation of the
LAT as well as scientific data analysis. These include the National Aeronautics and Space
Administration and the Department of Energy in the United States, the Commissariat \'a
l'Energie Atomique and the Centre National de la Recherche Scientifique/Institut National
de Physique Nucl\'eaire et de Physique des Particules in France, the Agenzia Spaziale Italiana
and the Istituto Nazionale di Fisica Nucleare in Italy, the Ministry of Education, Culture,
Sports, Science and Technology (MEXT), High Energy Accelerator Research Organization
(KEK) and Japan Aerospace Exploration Agency (JAXA) in Japan, and the K.A.Wallenberg
Foundation, the Swedish Research Council and the Swedish National Space Board in Sweden.

Additional support for science analysis during the operations phase is gratefully acknowledged
from the Istituto Nazionale di Astrofisica in Italy and the Centre National d'Etudes
Spatiales in France.

\bibliography{smithcloud} 

\end{document}

%% file: commands.tex
\newcommand{\ie}{i.e.\xspace}
\newcommand{\eg}{e.g.\xspace}

\mathchardef\mhyphen="2D

\newcommand{\vect}[1]{\boldsymbol{#1}}
\newcommand{\roughly}{\ensuremath{ {\sim}\,} }

\newcommand{\unit}[1]{\ensuremath{\mathrm{\,#1}}\xspace}
\newcommand{\MeV}{\unit{MeV}}
\newcommand{\GeV}{\unit{GeV}}
\newcommand{\TeV}{\unit{TeV}}
\newcommand{\degree}{\ensuremath{{}^{\circ}}\xspace}
\newcommand{\second}{\unit{s}}

\newcommand{\Myr}{\unit{Myr}}

\newcommand{\mm}{\unit{mm}}
\newcommand{\cm}{\unit{cm}}
\newcommand{\km}{\unit{km}}

\newcommand{\kpc}{\unit{kpc}}

\newcommand{\photon}{\unit{ph}}
\newcommand{\sr}{\unit{sr}}
\newcommand{\Kelvin}{\unit{K}}

\newcommand{\Msolar}{\ensuremath{\,M_\odot}}

\newcommand{\secref}[1]{Section~\ref{sec:#1}}

\newcommand{\tabref}[1]{Table~\ref{tab:#1}}

\newcommand{\Fermi}{\textit{Fermi}\xspace}

\newcommand{\TS}{\ensuremath{\mathrm{TS}}\xspace}

\newcommand{\code}[1]{\lstinline!#1!\xspace}

\newcommand{\GALPROP}{\code{GALPROP}}

\newcommand{\DM}{\ensuremath{\mathrm{DM}}}

\newcommand{\sigmav}{\ensuremath{\langle \sigma v \rangle}\xspace}

\newcommand{\PhiPP}{\ensuremath{\Phi_{\rm PP}}\xspace}

\newcommand{\bbbar}{\ensuremath{b \bar b}\xspace}

\newcommand{\ww}{\ensuremath{W^{+}W^{-}}\xspace}

\newcommand{\mumu}{\ensuremath{\mu^{+}\mu^{-}}\xspace}
\newcommand{\tautau}{\ensuremath{\tau^{+}\tau^{-}}\xspace}
\newcommand{\relic}{\ensuremath{3\times10^{-26}\cm^{3}\second^{-1}}\xspace}

\newcommand{\Jfactor}{\ensuremath{\mathrm{J\mhyphen factor}}\xspace}
\newcommand{\Jfactors}{\ensuremath{\mathrm{J\mhyphen factors}}\xspace}

\newcommand{\Mtidal}{\ensuremath{M_{\rm tidal}}\xspace}

\providecommand\physrep{\ref@jnl{Phys.~Rep.}}%
\providecommand\apjs{\ref@jnl{ApJS}}%
\providecommand{\jcap}{\ref@jnl{JCAP}}%
\providecommand{\plb}{\ref@jnl{PhLB}}%

\newcommand{\Hi}{\ion{H}{1}\xspace}
\newcommand{\NBH}{NBH09\xspace}
